\documentclass[prl,preprint,superscriptaddress]{revtex4-2}
\usepackage{graphicx}
\usepackage{hyperref}
\hypersetup{colorlinks = {true},linkcolor={blue},citecolor={blue},urlcolor={blue}} 
\usepackage[mathlines]{lineno}

\usepackage{xcolor}

\begin{document}
\preprint{}
\title{Real-time observation of frustrated ultrafast recovery from ionisation in nanostructured SiO$_2$ using laser driven accelerators.}
\author{J. P. Kennedy}%
\affiliation{Centre for Light-Matter Interactions, Department of Physics and Astronomy, Queen’s University Belfast, Belfast BT7 1NN, United Kingdom}
\author{M. Coughlan}%
\affiliation{Centre for Light-Matter Interactions, Department of Physics and Astronomy, Queen’s University Belfast, Belfast BT7 1NN, United Kingdom}
\author{C. R. J. Fitzpatrick}%
\affiliation{Centre for Light-Matter Interactions, Department of Physics and Astronomy, Queen’s University Belfast, Belfast BT7 1NN, United Kingdom}
\author{H. M. Huddleston}
\affiliation{Centre for Light-Matter Interactions, Department of Physics and Astronomy, Queen’s University Belfast, Belfast BT7 1NN, United Kingdom}
\author{J. Smyth}%
\affiliation{Centre for Light-Matter Interactions, Department of Physics and Astronomy, Queen’s University Belfast, Belfast BT7 1NN, United Kingdom}
\author{N. Breslin}%
\affiliation{Centre for Light-Matter Interactions, Department of Physics and Astronomy, Queen’s University Belfast, Belfast BT7 1NN, United Kingdom}
\author{H. Donnelly}%
\affiliation{Centre for Light-Matter Interactions, Department of Physics and Astronomy, Queen’s University Belfast, Belfast BT7 1NN, United Kingdom}
\author{C. Arthur}%
\affiliation{Centre for Light-Matter Interactions, Department of Physics and Astronomy, Queen’s University Belfast, Belfast BT7 1NN, United Kingdom}
\author{B. Villagomez}%
\affiliation{Centre for Light-Matter Interactions, Department of Physics and Astronomy, Queen’s University Belfast, Belfast BT7 1NN, United Kingdom}
\author{O. N. Rosmej}%
\affiliation{CGSI Helmholtzzentrum für Schwerionenforschung GmbH, Planckstr.1, 64291, Darmstadt, Germany}
\affiliation{Goethe University Frankfurt, Max-von-Laue-Str. 1, 60438, Frankfurt am Main, Germany}
\author{F. Currell}%
\affiliation{The Dalton Cumbria Facility and the School of Chemistry, The University of Manchester, Oxford Rd, Manchester M13 9PL, United Kingdom}
\author{L. Stella}%
\affiliation{Centre for Light-Matter Interactions, Department of Physics and Astronomy, Queen’s University Belfast, Belfast BT7 1NN, United Kingdom}
\affiliation{School of Chemistry and Chemical Engineering, Queen’s University Belfast, Belfast BT7 1NN, United Kingdom}
\author{D. Riley}%
\affiliation{Centre for Light-Matter Interactions, Department of Physics and Astronomy, Queen’s University Belfast, Belfast BT7 1NN, United Kingdom}
\author{M. Zepf}%
\affiliation{Helmholtz-Institute Jena, Fröbelstieg 3, 07743 Jena, Germany}
\affiliation{Friedrich-Schiller-Universität Jena, Physikalisch-Astronomische Fakultät, Max-Wien-Platz 1, 07743 Jena, Germany}
\author{M. Yeung}%
\affiliation{Centre for Light-Matter Interactions, Department of Physics and Astronomy, Queen’s University Belfast, Belfast BT7 1NN, United Kingdom}
\author{C. L. S. Lewis}%
\affiliation{Centre for Light-Matter Interactions, Department of Physics and Astronomy, Queen’s University Belfast, Belfast BT7 1NN, United Kingdom}
\author{B. Dromey}%
\email{b.dromey@qub.ac.uk}
\affiliation{Centre for Light-Matter Interactions, Department of Physics and Astronomy, Queen’s University Belfast, Belfast BT7 1NN, United Kingdom}

\date{\today}

\begin{abstract}
Ionising radiation interactions in matter can trigger a cascade of processes that underpin long-lived damage in the medium. To date, however, a lack of suitable methodologies has precluded our ability to understand the role that material nanostructure plays in this cascade. Here, we use transient photoabsorption to track the lifetime of free electrons ($\tau_c$) in bulk and nanostructured SiO$_2$ (aerogel) irradiated by picosecond-scale ($10^{-12}$ s) bursts of X-rays and protons from a laser-driven accelerator. Optical streaking reveals a sharp increase in $\tau_c$ from $<$ 1 ps to $>$ 50 ps over a narrow average density ($\rho_{av}$) range spanning the expected phonon-fracton crossover in aerogels. Numerical modelling suggests that this discontinuity can be understood by a quenching of rapid, phonon-assisted recovery in irradiated nanostructured SiO$_2$. This is shown to lead to an extended period of enhanced energy density in the excited electron population. Overall, these results open a direct route to tracking how low-level processes in complex systems can underpin macroscopically observed phenomena and, importantly, the conditions that permit them to emerge. 
\end{abstract}

\maketitle

As the size and complexity of a system grows, hierarchical sets of laws can emerge that define behaviour over different resolutions \cite{anderson1972}. An excellent illustration of this occurs in many-body physics. As the long-range order of atoms, or a lattice, is established it can spontaneously break the symmetry of free space leading to the emergence of phonons \cite{Vallone2020}. Gaining direct insight into how such a hierarchy can underpin a system's response is particularly relevant for radiation interactions in matter. This is because the evolution of deposited energy can span spatiotemporal scales ranging from \AA ($10^{-10}$~m) to $\sim$ cm and fs to $>$ ns and encompass everything from physical to chemical and biological processes \cite{anderson1972, Vallone2020, Ackland2010, Coughlan_2020, Mo2019, Kruijff2020, Cahill2014, Wingert_2016, Reid_1976, TOULEMONDE2012}. 

A common starting point for models tracking this transition from excitation to equilibrium is the mean free path between collisions $\lambda$: 
\begin{equation}
	\lambda = \frac{1}{n \sigma}
	\label{eq:1}
\end{equation}
where $n$ is number density and $\sigma$ is the interaction cross section. While Eq.\ref{eq:1} implies that $\lambda$ scales inversely proportional with $n$, the possibility of a non-uniform density \cite{Wingert_2016,Siddiqui2016, Orbach1986} calls into question the assumption of $\sigma$ being independent of $n$. For example, a heterogeneous distribution can have fundamentally different nanoscopic structure to that of a homogenous density, yet have the same average $n$ \cite{Wingert_2016, Orbach1986}. This leads to an unavoidable question -– how does the heterogeneity, or granularity, of a medium on the nanoscale affect the ultrafast processes that underpin recovery post-irradiation? As most media are not homogenous on these spatial scales (defects, disorder, grain boundaries etc), developing methods to test the impact of heterogeneity on $\sigma$ is central to our ability to realise a comprehensive model of nanoscopic energy transport in irradiated matter \cite{Cahill2014, Wingert_2016, Reid_1976, TOULEMONDE2012, Siddiqui2016, Orbach1986, surdutovich2019multiscale, tagliabue2020ultrafast, siemens2010quasi}. From engineering carrier lifetimes in nanostructured electronics deployed in radiation harsh environments \cite{Ackland2010, Coughlan_2020, Mo2019, TOULEMONDE2012} to investigating novel modalities for radiotherapy in healthcare \cite{Kruijff2020, surdutovich2019multiscale}, our ability to transform existing technologies will rely on predicting and controlling \cite{Ackland2010} the evolution of dose on the shortest spatial and fastest temporal scales post-irradiation.

\begin{figure*}
\centering
	\includegraphics[width=0.92\textwidth]{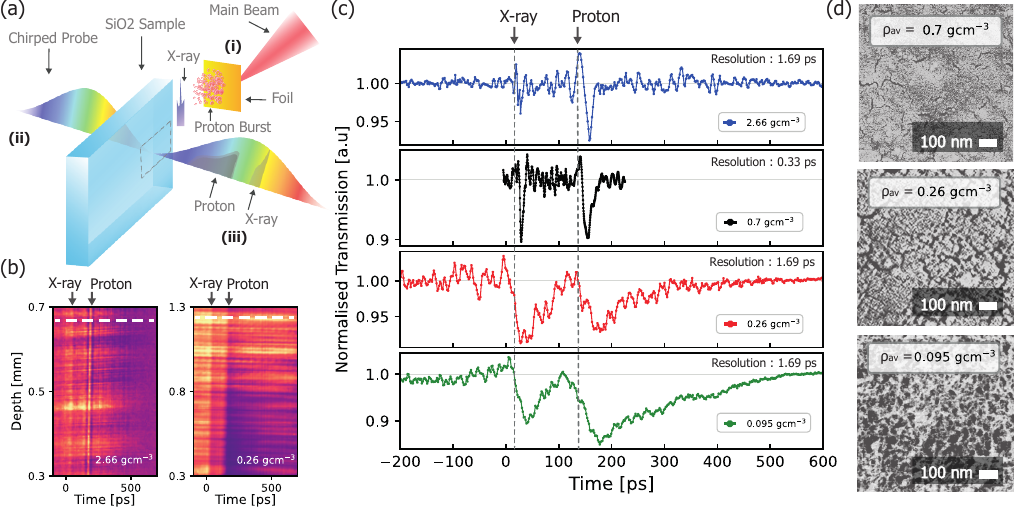}
	\caption{Spatiotemporally resolved ionisation dynamics for picosecond scale pulses of X-rays and protons stopping in bulk and nanostructured SiO$_2$. Sketch \textbf{(a)} shows the experimental setup. \textbf{(i)} The main laser is incident onto a foil target generating a bright broadband pulse of X-ray radiation and burst of protons via the TNSA mechanism \cite{Macchi2013}. \textbf{(ii)} A linearly chirped probe synchronised with the main laser is incident onto a sample. \textbf{(iii)} As the chirped probe passes through the sample, the ionisation dynamics due to the radiation generated by the main laser are encoded temporally in the observed spectrum of the probe. In \textbf{(b)} experimentally obtained optical streak data for opacity induced in SiO$_2$ with average densities ($\rho_{av}$) $2.66$~gcm$^{-3}$ (bulk) and $0.26$~gcm$^{-3}$ (aerogel) are shown. Plot \textbf{(c)} shows the background subtracted lineouts for interactions in $2.66$~gcm$^{-3}$ (bulk, blue dots), $0.7$~gcm$^{-3}$ (xerogel, black dots), $0.26$~gcm$^{-3}$ (aerogel, red dots) and $0.095$~gcm$^{-3}$ (aerogel, green dots) taken from the raw optical streak data (shown in \textbf{(b)}) at the white dashed lines. Data for each sample is obtained on a single shot with its corresponding resolution shown on each plot. Time is given with respect to the time of the laser interaction with the foil target. The depths where lineouts are taken are chosen such that the absolute proton pulse duration is at a constant to allow direct comparison (See Supplemental Material \cite{supp}). Scanning electron microscope images for samples of various densities are shown in  \textbf{(d)}. The magnification is 33,000 times, and the scale bar of 100~nm, is the same for all images.}
	\label{fig.2}
\end{figure*} 

In general, multiple models based on $\lambda$ are cascaded to interpret this dynamic phase. The initial dose distribution is obtained from the linear energy transfer (or stopping power) of the irradiating species \cite{HENKE1993, Ziegler_1999}. Next, thermal spike model \cite{TOULEMONDE2012,klaumunzer2006,kaganov1957} can be invoked for the subsequent energy density evolution and thermalisation with the background material. A key element for describing these dynamics is the role of collective medium response. This can be approximated using the phonon gas model (PGM) which assumes a quasi-infinite periodic lattice to support delocalised plane-wave vibrational modes, or phonons \cite{Vallone2020, Reid_1976, lv2016examining}. At the same time, for disordered media, discrepancies between PGM predictions and observations for thermal conductivities and vibrational spectra are addressed by invoking phonon localisation \cite{Anderson1958, Graebner1986, Luckyanova2018}. 

Unfortunately, these observables typically encompass a broad range of processes averaged over macroscopic regions. This makes their use in understanding how localisation affects material response and recovery to ionising radiation a challenge due to the nascent action of ionising species in matter occurring on ultrafast and nanoscopic spatiotemporal scales. To overcome this challenge we track recovery time (free carrier lifetime), $\tau_c$, in bulk and nanostructured SiO$_2$ irradiated by ps pulses of X-rays and protons. As they transfer their energy, these ionising species generate electron spectra with distinctly different average energies that, as a result, interrogate the material response over distinctly different nm length scales. This duality provides a unique tool with which to study how localisation of vibrational modes impacts $\tau_c$ and, ultimately, represents a test of how changing complexity on the nanoscale affects macroscopic response.

When a relativistically intense laser pulse ($>10^{18}$~Wcm$^{-2}$) is incident at an oblique angle (30$^\circ$ - 45$^\circ$) onto a $\mu$m-scale thick metal foil target, electrons from the front surface are driven through the foil by the strong electric field of the driving laser. This provides the basis for two highly synchronous ultrafast bursts of radiation from a single high-power laser pulse - a bright broadband pulse of Bremsstrahlung X-ray radiation and burst of protons via the Target Normal Sheath Acceleration (TNSA) mechanism \cite{Macchi2013}. While Bremsstrahlung from laser matter interactions is a well-established source of $> 10$~keV X-rays with ps scale duration, only recently have experiments verified that TNSA pulse durations as short as $\sim 3$~ps are possible from the same interactions \cite{dromey2016}. Monte Carlo modelling performed using Geant4 \cite{AGOSTINELLI2003} shows that protons ($\leq 10$~MeV) generate dense nanotracks of ionisation with average electron energy, $E_{av}$, $< 100$~eV. Conversely, X-rays ($>10$~keV, $<100$~keV) undergo multiple high momentum transfer scattering events that homogenise the initial dose distribution and result in $E_{av}$ on the order of $25$~keV (See Supplemental Material \cite{supp}). In Fig.\ref{fig.2}(c) we track $\tau_c$ following the interaction of these ps-scale pulses in bulk and nanostructured SiO$_2$ with a range of average densities ($\rho_{av}$) via single-shot optical streaking using the laser-driven accelerator at the TARANIS facility in Queen's University Belfast \cite{dzelzainis_2010, dromey2016}. Silica aerogels \cite{Schaefer1986, Courtens1987, anglaret1994, nakayama2003fractal, Gefen1983, Caponi2004, courtens1988observation} were used which comprise of a nanostructured matrix of solid density SiO$_2$ nanoparticles with an average size of $a = 3 \pm 1$~nm and tuneable average pore size (See Fig.\ref{fig.2}(d)). These nanostructured media can be described as a percolating fractal network across spatial scales from $l_{F}$ up to an acoustic correlation length $\xi_{ac}$ (where $a < l_{F} < \xi_{ac}$) \cite{nakayama2003fractal, Gefen1983, Caponi2004}. The acoustic correlation length $\xi_{ac}[\AA]$ can be estimated as $8.3 \times 10^{6} \times \rho_{av[\text{kg} \cdot \text{m}^{-3}]}^{-1.84}$ \cite{courtens1988observation}. In SiO$_2$, Longitudinal Optical (LO) phonon emission \cite{SCHREIBER2002} and polaronic stopping (POL) \cite{STEPANOV1995, Dapor_2012} represent the primary mechanisms of energy loss for electrons with average energies ($E_{av}$) within the sub-excitation range in SiO$_2$. As such, aerogels provide control over heterogeneity on spatial scales relevant for scattering of sub-excitation energy electrons from LO phonons ($\lambda_{LO}$) in SiO$_2$ \cite{SCHREIBER2002}. 

Optical streaks obtained for X-rays followed by slower protons interacting in bulk and a range of nanostructured SiO$_2$ samples are shown in Fig.\ref{fig.2}. These materials and densities straddle the transition expected for fractal behaviour in nanostructured SiO$_2$ \cite{courtens1988observation}. As can be seen there is a sharp increase in $\tau_c$ for both the X-rays and protons interacting in the 0.26 gcm$^{-3}$ aerogel sample. Based on the simplistic $\tau_c \propto \rho_{av}^{-1}$ scaling, one would expect a moderate $< 3$ times increase in $\tau_c$. Instead, we observe factors of $50 - 70$ times longer in $\tau_c$ for both X-rays and protons interacting in aerogel. 
As mentioned above, $E_{av}$ for electrons due to $10$~MeV proton stopping in SiO$_2$ is on the order 10's of eV. From Fig.\href{URL_will_be_inserted_ by_publisher}{SM 2.1} in \cite{supp} (based on \cite{SCHREIBER2002}), $\lambda$ for these electrons in bulk SiO$_2$ is $<3$~nm. The implication is that the dose due to $10$~MeV proton stopping directly interrogates the nanoscopic heterogeneity in aerogel, which is at bulk density. In contrast, $E_{av}$ due to a bremsstrahlung source with a $50$~keV temp is 10's of keV. From \cite{supp}, $\lambda$ for these electrons in bulk SiO$_2$ is $> 100$~nm and considerably larger in the lower density samples (dashed line in Fig.\href{URL_will_be_inserted_ by_publisher}{SM 2.1}). This is significantly larger than $\xi_{ac}$ predicted by the scaling from \cite{courtens1988observation}. This implies that these electrons probe the medium over lengths scales on which homogeneous behaviour is expected. 

\begin{figure}
	\includegraphics[width=0.6\textwidth]{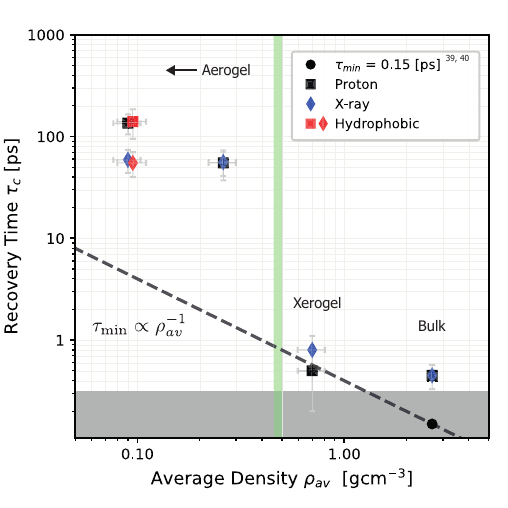}
	\centering
	\caption{Dependence of recovery time $\tau_c$ on average density ($\rho_{av}$). As the nanostructure changes from bulk to percolating fractal network a distinct discontinuity in the scaling of $\tau_c$ with respect to $\rho_{av}$ is observed. The presence of H$_2$O as a potential cause for the observed increase in $\tau_c$ is removed by using an aerogel sample with hydrophobic surface chemistry (red symbols, density $0.1$~gcm$^{-3}$). The green shaded region outlines the expected boundary between bulk and fractal phenomenology. The black dashed line shows the expected scaling assuming a homogenous reduction in $\rho_{av}$($\tau_{ac} \propto \rho_{av}^{-1}$) and is based on the experimentally observed $\tau_{\text{min}}$ = 0.15~ps for $\rho_{av}$ = 2.66~gcm$^{-3}$ from Audebert \textit{et al}. \cite{Audebert1994}. The grey shaded region is the minimum resolution for the optical streaking performed here ($0.33 \pm 0.05$~ps).}
	\label{fig.3}
\end{figure}

Fig.\ref{fig.3} shows dependence of the recovery time on average density ($\tau_c \propto \rho_{av}^{-1}$) which reveals a sharp discontinuity in the scaling. As the material structure changes from root mean square (rms) pore size $< 5$~nm to rms pore size $>10$~nm (Fig.\ref{fig.2}(d)) the ultrafast response in SiO$_2$ due to self-trapped exciton formation (STE) \cite{Audebert1994} is observed to quench, with $\tau_c <$~ps growing to $\tau_c > 50$~ps. Importantly, this switching is observed for recovery due to irradiation from both X-rays ($\xi_{ac}$) and protons ($l_f$). While this is a strong indication that ultrafast STE formation is being inhibited in aerogel, it is important to note that this does not preclude STE formation in general. It simply implies that the time taken to meet the trapping criterion put forward by Martin \textit{et al}. \cite{Martin1997} is significantly longer than would be expected if the $\tau_c \propto \rho_{av}^{-1}$ scaling is applied to the result of Audebert \textit{et al}. \cite{Audebert1994} (grey dashed line, Fig.\ref{fig.3}).

\begin{figure}
	\includegraphics[width=0.6\textwidth]{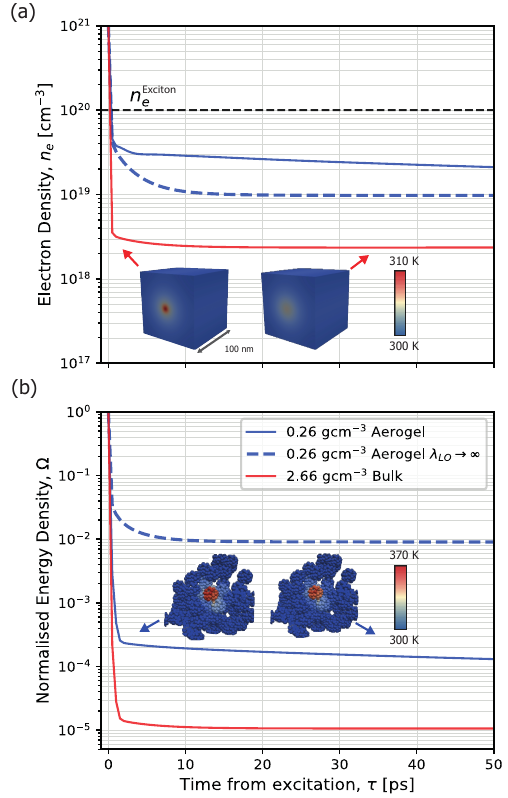}
	\caption{Hydrodynamic modelling of the hot electron population evolution. In \textbf{(a)} the evolution of the electron density following the interaction of a single proton in bulk and aerogel samples demonstrates a significant increase in electron density over extended time frames. The corresponding normalised energy density is shown in \textbf{(b)} (solid red and blue lines in both figures, respectively), assuming $\lambda \propto \rho_{av}^{-1}$. Conversely, if $\lambda_{\text{LO}} \rightarrow \infty$ is set for aerogel the dashed traces are obtained for the calculations. The figure insets show lattice temperature due to the deposited dose evolves spatially for bulk and nanostructured SiO$_2$. In both cases, the maximum lattice temperature is significantly lower than both melting ($>1700$~K) and Debye temperature ($470$~K) in SiO$_2$.}
	\label{fig.4}
\end{figure}

To help interpret these observations, we implement a energy transfer model \cite{kaganov1957, klaumunzer2006, grasser2003review} to track the early stage evolution of the thermal electron population caused by a single proton stopping in bulk SiO$_2$ and aerogel with $\rho_{av} = 0.26$~gcm$^{-3}$. First, we consider the situation where the cross sections for interactions are independent of structure on the nanoscale. This is given by the solid traces in Fig.\ref{fig.4}(a) and \ref{fig.4}(b), with red for bulk and blue for nanostructured SiO$_2$

Fig.\ref{fig.4}(a) shows that following a rapid diffusion phase ($<0.5$~ps), there is a clear effect of the nanostructure in slowing the evolution of electron density, $n_e$, in aerogel compared to that in bulk SiO$_2$, with $n_e(\text{bulk}) < n_e(\text{aerogel})$ for $\tau$ extending to $50$~ps. However, for $\tau < 0.2$~ps in both materials, $n_e$ is below $n_e^{\text{Exciton}} \approx 10^{20}$ cm$^{-3}$. This is the maximum free carrier density observed for STE formation in SiO$_2$ from experiments \cite{mauclair2016excitation}. Accordingly, confinement of a near-solid density population of hot electrons in the nanostructure can be ruled out as the source of enhanced $\tau_c$ in aerogel. Similarly, Fig.\ref{fig.4}(b) reveals a corresponding increase in the kinetic energy density ($\Omega$) for the electron population in nanostructured SiO$_2$. This increase is largely reconciled by the increase in $n_e$. Therefore, for these conditions, similar $E_{av}$ is expected for electrons ionised by protons stopping in both bulk and nanostructured SiO$_2$. This suggests that the trapping criterion of Martin \textit{et al}. \cite{Martin1997} should not be violated in aerogel for conditions where ultrafast trapping is observed in the bulk sample. Our initial assumption of constant $\sigma$ fails to expain for the enhanced $\tau_c$ observed in aerogel. To reconcile this discrepancy, we test how the nanoscopic heterogeneity of the aerogel may lead to the suppression of loss mechanisms for sub-excitation energy electrons. Assuming the cross section for LO phonon emission ($\sigma_{\text{LO}}^{-}$) becomes vanishingly small, i.e. ${\lambda_{\text{LO}} \rightarrow \infty}$ the above picture changes considerably. For $\tau > 10$~ps our model suggests $> 10^{2}$ times increase in $\Omega$ over that expected in bulk SiO$_2$ (blue dash line, Fig.\ref{fig.4}(b)). This implies that the requirement for $E_{av}$ to approach the thermal energy of the lattice for trapping via excitonic states is no longer met on ultrafast ($< 1$~ps) timescales.

The assumption that $\lambda_{\text{LO}} \rightarrow \infty$ in aerogel is supported in several ways. Firstly, it does not suggest that vibrations cease to exist in aerogel. The hierarchy of fractons and complex surface modes of the nanoparticles comprising the aerogel substructure can play a key role in the carrier dynamics post irradiation \cite{Schaefer1986, Courtens1987, anglaret1994, nakayama2003fractal, Ramayya2008}. However, when it is considered that electrons with $E_{av} < 0.9$~eV (electron affinity in SiO$_2$) \cite{STEPANOV1995} are essentially confined in the nanoscopic network (i.e. Fig.\ref{fig.4}(a)) this can lead to a non-equilibrium phonon distribution due to a locally enhanced $n_e$. For these conditions the cross section for reabsorption of optical phonons ($\sigma_{\text{LO}}^{+}$) approaches $\sigma_{\text{LO}}^{-}$ approximating to $\lambda_{\text{LO}} \rightarrow \infty$ (hot phonon bottleneck) \cite{yang2016observation}. Secondly, LO phonons decay into two acoustic phonons on $1$~ps timescales \cite{Klemens1966}. While long-wavelength acoustic modes ($>\xi_{ac}$) can still propagate in aerogel, albeit at considerably reduced velocity to their bulk counterparts \cite{Caponi2004}, recent work has begun to question the relative stopping powers for optical and short-wavelength acoustic phonons in bulk materials \cite{Fischetti2019}. In aerogel these short wavelength acoustic modes will manifest as fractons with short localisation length for $l_F$ ($< \xi_{ac}$). This implies a significantly reduced spatial overlap with free carriers in aerogel over bulk, subsequently reducing the probability of energy loss \cite{Orbach1986}. Finally, it is also important to recognise the significance of the quasi-infinite periodic lattice assumed in the PGM. As outlined earlier, the periodicity of the lattice spontaneously breaks the Galilean symmetries of free space leading to the emergence of Higgs and Goldstone modes, or optical and acoustic phonons respectively \cite{Vallone2020}. Conversely, the nanoscopic self-similarity of fractal networks implies that these systems do not possess the long-range translational symmetry of a crystal lattice i.e. the assumption of quasi-infinite periodicity is invalid. In this sense, Galilean symmetries are not spontaneously broken in self-similar/fractal materials as they are not translationally invariant \cite{anglaret1994, nakayama2003fractal} and the assumption that $\lambda_{\text{LO}}\rightarrow \infty$ in aerogels can be grounded in basic symmetry arguments. Although the modelling in Fig.\ref{fig.4} provides insight into the excited electron population dynamics over short distances, the rigorous analysis required to fully interrogate recombination and trapping dynamics for the excited electron population in a numerical framework for this phase is beyond the scope of the work here \cite{stella2021excitonic}.  

In conclusion, this work further supports the idea that the fundamental nature of atomic scale vibrations, and what may be termed coherent phonon modes must be considered carefully in the context of disordered and nanostructured materials. Our work shows that this is particularly relevant for understanding the evolution of deposited energy due to ionising radiation in such media. The observation of a sharp discontinuity in the scaling of $\tau_c$ for a heterogeneous reduction in $\rho_{av}$ demonstrates that the efficient mechanisms for ultrafast relaxation of sub-excitation energy electrons via LO phonon emission is not dominant in aerogels over length scales where both strongly localised ($l_{f} < \xi_{ac}$) and delocalised ($L > \xi_{ac}$) phonon modes are expected. These length scales are probed by primary electrons due to proton and X-ray stopping respectively. This leads to the conclusion that ultrafast decay of free carriers in SiO$_2$ emerges as the nanoscale structure of the material is modified to that of a non-fractal medium with long-range ($>5$~nm) periodic lattice. Furthermore, this work demonstrates how the multi-species capabilities of laser-driven accelerators can provide a platform for future investigations on the role of anomalous diffusion on fractal networks and in disordered systems in real time. It also provides the potential for highly synchronised harmonic probe beams with attosecond resolution \cite{seiffert2017attosecond} for investigating the fundamental nature of energy loss via electron-phonon interactions (i.e. LO versus acoustic modes) \cite{Fischetti2019} in irradiated materials on the primary excitation timescales.

\begin{acknowledgments}
We would like to thank V. P. Efremov for supplying aerogel samples. Work is funded through EPSCR grants (EP/P010059/1, EP/P016960/1 and EP/W017245/1). We are also grateful for use of the computing resources from the Northern Ireland High Performance Computing (NI-HPC) service funded by EPSRC (EP/T022175). L.S. also acknowledges funding from the European Union's Horizon 2020 research and innovation programme under the Marie Sk\l odowska-Curie Actions staff exchange Grant agreement No. 823897 (MSCA-RISE "ATLANTIC"). L.S. and J.S. also  acknowledgments support from the COST Action CA17126 'TUMIEE'. 
\end{acknowledgments}

\bibliographystyle{apsrev}
\bibliography{arerogel_bib.bib}

\end{document}